\documentclass[lettersize,journal]{IEEEtran}
\usepackage{amsmath,amsfonts}
\usepackage{algorithmic}
\usepackage{algorithm}
\usepackage{array}
\usepackage[caption=false,font=normalsize,labelfont=sf,textfont=sf]{subfig}
\usepackage{textcomp}
\usepackage{stfloats}
\usepackage{url}
\usepackage{verbatim}
\usepackage{graphicx}
\usepackage{cite}
\usepackage{graphicx}
\usepackage{adjustbox}
\usepackage{tabularx}
\usepackage{booktabs}
\usepackage{multirow}
\usepackage{amsmath}
\usepackage{float}
\usepackage{amssymb}
\usepackage{xcolor}
\usepackage{hyperref}
\begin{document}

\title{FPT-Noise: Dynamic Scene-Aware Counterattack for Test-Time Adversarial Defense in Vision-Language Models}

\author{
    \IEEEauthorblockN{Jia Deng$^{1}$, Jin Li$^{1}$, Zhenhua Zhao$^{2}$, Shaowei Wang$^{1*}$\thanks{Corresponding Author. Email: wangsw@gzhu.edu.cn}}
    
    \IEEEauthorblockA{$^{1}$ School of Artificial Intelligence, Guangzhou University, Guangzhou, China}
    
    \IEEEauthorblockA{$^{2}$ ByteDance, Beijing, China}
}


\markboth{Journal of \LaTeX\ Class Files,~Vol.~14, No.~8, August~2021}%
{Shell \MakeLowercase{\textit{et al.}}: A Sample Article Using IEEEtran.cls for IEEE Journals}


\maketitle

\begin{abstract}
Vision-Language Models (VLMs), such as CLIP, have demonstrated remarkable zero-shot generalizability across diverse downstream tasks. However, recent studies have revealed that VLMs, including CLIP, are highly vulnerable to adversarial attacks, particularly on their visual modality.  Traditional methods for improving adversarial robustness, such as adversarial training, involve extensive retraining and can be computationally expensive. In this paper,  we propose a new Test-Time defense: Feature Perception Threshold Counterattack Noise (FPT-Noise), which enhances the adversarial robustness of CLIP without costly fine-tuning. Our core contributions are threefold: First, we introduce a Dynamic Feature Modulator that dynamically generate an image-specific and attack-adaptive noise intensity parameter. Second, We reanalyzed the image features of CLIP. When images are exposed to different levels of noise, clean images and adversarial images exhibit distinct rates of feature change. We established a feature perception threshold to distinguish clean images from attacked ones. Finally, we integrate a Scene-Aware Regulation guided by a stability threshold and leverage Test-Time Transformation Ensembling (TTE) to further mitigate the impact of residual noise and enhance robustness.Extensive experimentation has demonstrated that FPT-Noise significantly outperforms existing Test-Time defense methods, boosting average robust accuracy from 0.07\% to 56.86\% under AutoAttack while maintaining high performance on clean images (–1.1\%). The code will be made public following the publication of the study. The code will be made public following the publication of the study. 

\end{abstract}


\section{Introduction}
Vision-Language Models (VLMs) have emerged as powerful tools across a wide range of applications, including image classification, visual question answering, and image captioning. These models leverage large-scale image-text pair datasets for pre-training, among which CLIP \cite{clip} demonstrates remarkable zero-shot generalizability across downstream tasks—setting it apart in this domain. CLIP’s zero-shot classification capability enables effective image categorization without relying on extensive labeled data, thereby reducing annotation costs and facilitating rapid adaptation to new tasks. Additionally, the model’s pre-training integrates rich visual and linguistic information from large-scale datasets, laying a robust foundation for its strong performance in diverse image classification tasks.

\begin{figure}[!tb]
    \centering
    \includegraphics[width=\linewidth]{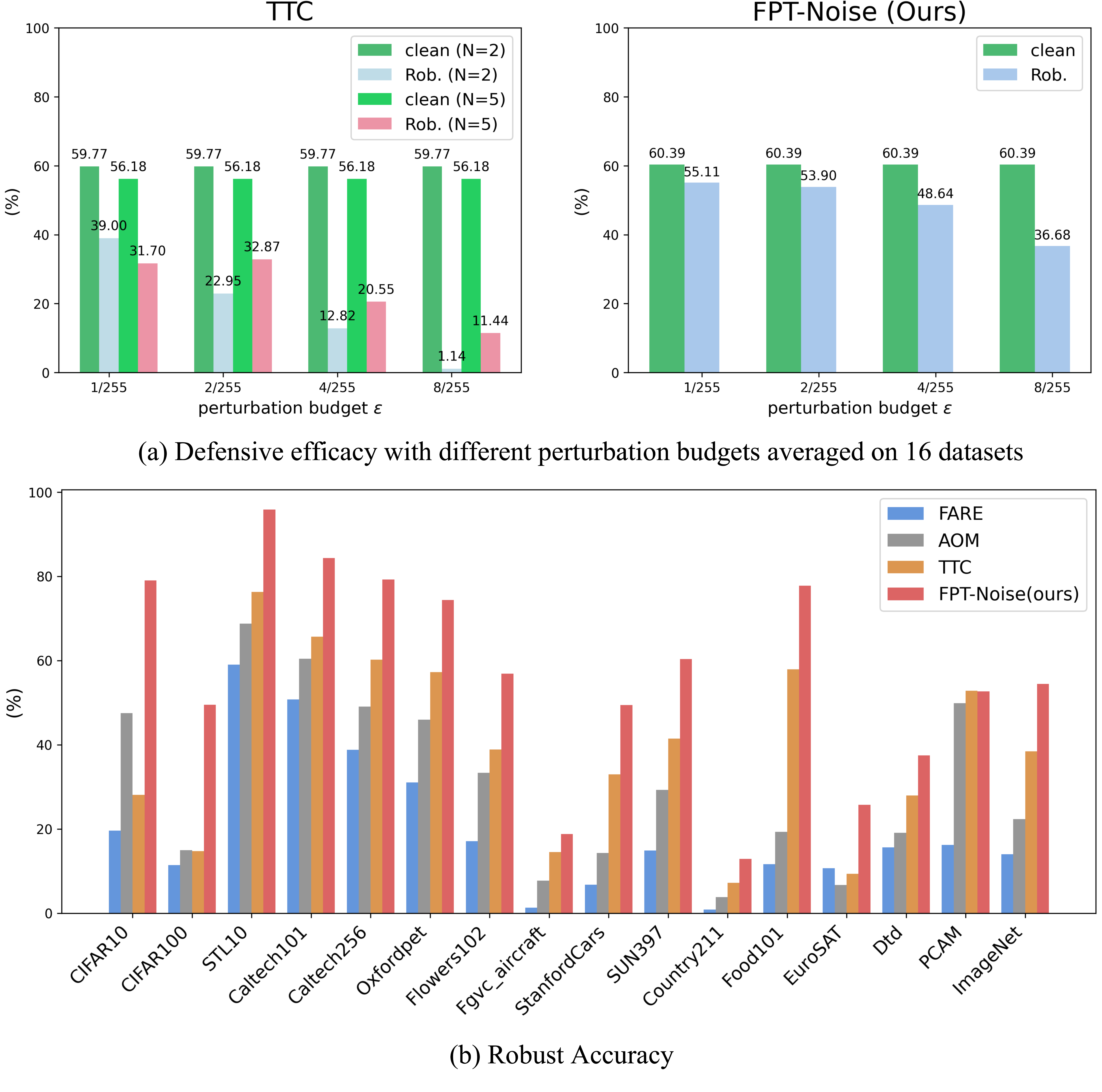}
\caption{(a) We tested the defensive efficacy of TTC and FPT-Noise against PGD-10 with different perturbation budgets averaged on 16 datasets. (b) Classification accuracy (\%) on both adversarial images (Robs.) under PGD-10 attack at $\epsilon=1/255$.}

    \label{fig1}
\end{figure}

However, recent studies have highlighted the vulnerability of VLMs (including CLIP) to adversarial attacks, with their visual modality being particularly susceptible. Even minute, imperceptible perturbations can significantly degrade model performance: for instance, under AutoAttack, CLIP’s average robust accuracy on adversarial images drops to a mere 0.07\% \ref{t2}—posing substantial safety risks in real-world applications such as autonomous driving and medical image diagnosis. The present study focuses on enhancing CLIP’s adversarial robustness against imperceptible perturbations injected into input images, while preserving its core advantage of zero-shot generalization.

Traditional methods for enhancing adversarial robustness (e.g., adversarial training) involve augmenting training data with adversarial examples and retraining models to resist perturbations. While effective, these approaches require extensive retraining and incur substantial computational costs—especially for large-scale models like CLIP. For models with high training costs, most methods rely on fine-tuning: (1) Adversarial Fine-Tuning (AFT), which alternates between generating adversarial images on a single dataset and fine-tuning CLIP’s vision encoder; (2) Adversarial Prompt Tuning (APT), which aligns learnable text prompts with adversarial image embeddings to provide a more cost-efficient alternative. Nevertheless, even with fine-tuning, generating adversarial images remains time-consuming, and fine-tuning on a single dataset often leads to overfitting—diminishing zero-shot generalization capabilities.

Recent studies have explored test-time defense strategies to avoid retraining/fine-tuning costs. For example, Wu et al. \cite{wu2021attacking} demonstrated that adding random noise to adversarial examples in adversarially trained models can negate misleading predictions, but their fixed-noise scheme fails to adapt to varying image complexities. Xing et al. \cite{xing2025clip} proposed a test-time paradigm (TTC) that leverages CLIP’s expressiveness for self-defense, noting that adversarial images exhibit higher stability than clean images when small random noise is introduced; however, TTC suffers from two critical flaws: it requires manual adjustment of counterattack steps for different attack strengths (e.g., 2 steps for \(\epsilon=1/255\) vs. 5 steps for \(\epsilon=4/255\), as shown in Figure \ref{fig1}(a)), which introduces deployment costs and failure risks due to suboptimal parameter tuning, and its fixed-intensity counter-noise cannot adapt to image-specific features (e.g., simple MNIST digits vs. complex ImageNet objects). Tong et al. \cite{aom} proposed a training-free zero-shot method (AOM) that constructs a linear path in the embedding space to shift the adversarial domain toward a cleaner domain, but it cannot distinguish between clean and adversarial samples; applying its interpolation strategy to clean images causes unnecessary performance degradation (e.g., a 3–4\% drop in clean accuracy on CIFAR-10, as shown in Table \ref{t1}).

These observations lead to three fundamental research questions that existing methods fail to address comprehensively:
\begin{itemize}
\item \textbf{Adaptivity}: How can we dynamically generate counter-noise that is both image-specific and attack-adaptive, surpassing fixed-intensity schemes?
\item \textbf{Discrimination}: How can we accurately distinguish adversarial samples from clean ones at test time to avoid unnecessary processing?
\item \textbf{Generalization}: How can we achieve this without manual parameter tuning for different datasets or attacks, preserving true zero-shot capability?
\end{itemize}

To address these questions, we propose \textbf{FPT-Noise} (Feature-Perception Threshold Counterattack Noise)—not merely an incremental enhancement, but a holistic, integrated framework that redefines the test-time defense paradigm for VLMs. Our solution is built on three synergistic pillars: (1) a Dynamic Feature Modulator (DFM) that intelligently calculates the optimal counter-noise intensity in real time; (2) a Feature Perception Threshold (FPT): a novel metric rooted in the differential feature-space behavior of clean and adversarial images under noise, which enables accurate sample discrimination; (3) a Scene-Aware Regulation (SAR) module that orchestrates targeted countermeasures based on the FPT’s output, ensuring clean samples remain largely unperturbed. 

Experiments across 16 benchmark datasets demonstrate that FPT-Noise significantly outperforms existing test-time defense methods: under AutoAttack (\(\epsilon=1/255\)), it boosts CLIP’s average robust accuracy from 0.07\% to 56.86\%, while maintaining high clean accuracy (only a 0.77\% average drop, as shown in Table \ref{t2}). By eliminating the need for fine-tuning and manual parameter adjustment, FPT-Noise balances robustness, efficiency, and zero-shot generalization—making it suitable for real-world deployment.



\section{Related Work}
\subsection{Adversarial attack and defense}
Adversarial noise robustness is a critical challenge in deep learning  \cite{19,16}. Adversarial attacks fall into black-box, white-box, and adaptive frameworks: black-box attacks use zeroth-order optimization to maximize losses  \cite{uesato2018adversarial,guo2019simple}; white-box attacks  \cite{FGSM,BIM,PGD} exploit full network knowledge. Examples like Carlini-Wagner (CW)  \cite{CW} and AutoAttack (AA)  \cite{AA} create imperceptible gradient-based perturbations, degrading image/text accuracy  \cite{text,text2}.

Adversarial Training (AT)  \cite{AT} remains a key defense, enhancing robustness via training-time parameter regularization and perturbation integration. However, AT suffers from long training times, high data demands, and a known robustness-accuracy trade-off  \cite{tsipras2018robustness}.  A prevalent paradigm of adversarial purification aims to maximize the log-likelihood of samples to remove perturbations in pixel space. Diffusion-based adversarial purification methods, such as those  \cite{nie2022diffusion, chen2023robust, zhang2024classifier}, have been shown to achieve state-of-the-art robustness. However, the processing of a single image is a time-consuming process.
To address this, Test-Time defenses have emerged: data augmentation  \cite{perez2021enhancing,zhang2021memo,shanmugam2021better} improves robustness via input adjustments; purification techniques use generative models  \cite{yoon2021adversarial,nie2022diffusion} or pre-trained CNNs  \cite{wang2023addition,zhang2024detecting} to remove noise. Our approach aligns with Test-Time counterattacks, which inject adaptive noise to restore model correctness.

\subsection{Adversarial robustness of CLIP}

The rising popularity of large visual-language models has heightened interest in their vulnerability to adversarial attacks, with CLIP—a key model for image-text alignment—being a focus for enhancing adversarial robustness. Mao et al. \cite{mao2022understanding} introduced TeCoA, which fine-tunes CLIP's vision encoder using real-time adversarial samples on a single dataset to transfer robustness to downstream tasks. Wang et al. \cite{wang2024pre} enhanced this by using the original CLIP to guide adversarial fine-tuning with two regularization terms, boosting generalization on both adversarial and clean images. Schlarmann et al. \cite{schlarmann2024robust} proposed FARE, an unsupervised fine-tuning scheme for CLIP's vision embeddings to enhance robustness while preserving original features. Xing et al. \cite{xing2025clip} developed a test-time paradigm leveraging CLIP's expressiveness for defense, noting adversarial images are more stable under small random noise. Tong et al. \cite{aom}  constructs a linear path in the embedding space to interpolate and move the adversarial domain into a cleaner domain. Wang et al. \cite{wang2025tapt} introduced TAPT to enhance inference robustness, and Zhang et al. \cite{zhang2025clipure} proposed CLIPure, a latent-space purification method for adversarially robust zero-shot classification. 

While existing methods for enhancing CLIP's adversarial robustness either perform poorly, are time-consuming (e.g., adversarial training with extensive retraining, Test-Time defenses like TTC requiring manual parameter tuning), our proposed FPT-Noise achieves significantly better performance efficiently without costly fine-tuning.

\begin{figure*}
    \centering
    \includegraphics[width=0.98\linewidth]{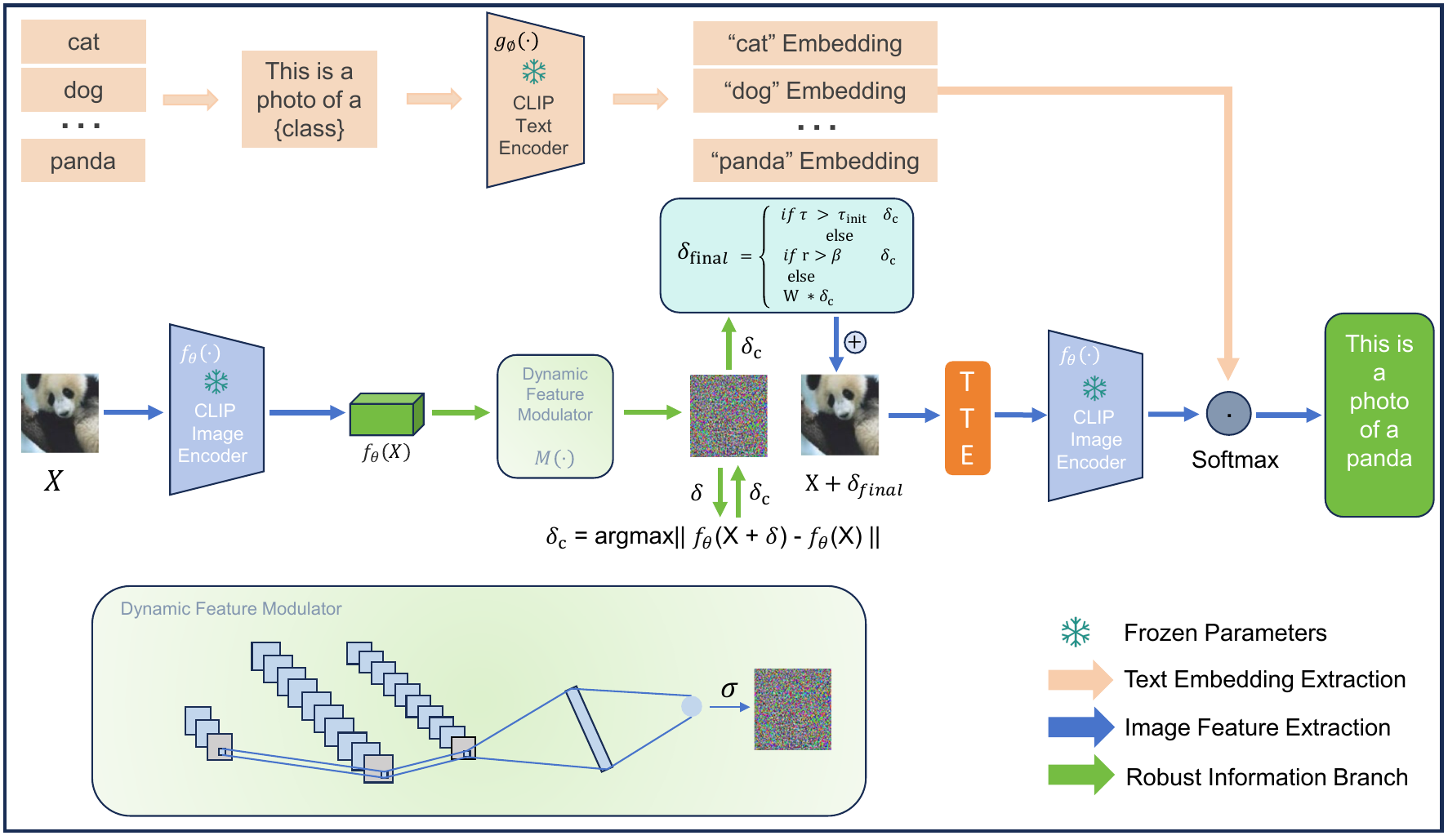}
    \caption{The pipeline of FPT-Noise.  Pre-trained CLIP acquires text embeddings and image features $f_\theta(X)$, which are processed by the Dynamic Feature Modulator ($M$). The counterattack noise, denoted by $\delta_{c}$, is generated via the loss function. Using our designed Feature Perception Threshold and Scene-Aware Regulation, control the intensity of the noise, then add it to the image and then TTE is performed. The Softmax function, when integrated with text embeddings, yields the desired output.}

    \label{pipe}
\end{figure*}
\section{Preliminaries}

\subsection{Zero-shot Classification of CLIP}

CLIP (Contrastive Language–Image Pre-training) \cite{clip} is a deep learning model that aligns images and texts by embedding them into a shared feature space. It has demonstrated remarkable performance in zero-shot classification tasks, thereby showcasing its ability to learn without direct supervision. Specifically, zero-shot classification is achieved by leveraging CLIP's pre-trained vision encoder, denoted by \( f_\theta(\cdot) \), and its text encoder, denoted by \( g_\phi(\cdot) \), to align images and text. In this context, a test image, denoted by \( X \), and a set of class names, denoted by \( c_1, c_2, \ldots, c_K \), are provided as inputs. CLIP employs a cosine similarity computation to classify the image by measuring the similarity between the image embedding and the text prompt embedding. The cosine similarity \( s_i \) between the image embedding \( f_\theta(X) \) and the text prompt embedding \( g_\phi(T(c_i)) \) is calculated as follows:

\begin{equation}
    s_i = \frac{f_{\theta}(X) \cdot g_{\phi}(T(c_i))}{\| f_{\theta}(X) \| \cdot \| g_{\phi}(T(c_i)) \|}
\end{equation}

Here, \( T(c_i) \) represents a text prompt for class \( c_i \) (e.g., "This is a photo of a [CLASS]"). The image \( X \) is classified as the class with the highest similarity \( s_i \).

\subsection{Adversarial Attacks for CLIP}

Despite CLIP's impressive performance in image matching tasks, recent studies have highlighted its vulnerability to adversarial attacks. Adversarial attacks involve adding small, imperceptible perturbations to input images with the intention of causing incorrect model outputs. This type of attack poses a significant threat to the robustness of many deep learning models, including CLIP.

The objective of adversarial attacks is to add a perturbation \( \delta \) to the image \( x \) such that the classification loss function \( L \) of CLIP is maximized, while ensuring that the perturbation remains imperceptible to humans. Mathematically, this can be formulated as:
\begin{equation}
    \delta_a = \arg \max_{\delta} L(x + \delta, t_c), \quad \text{subject to} \quad \| \delta \|_p \leq \epsilon_a
\end{equation}

Here, \( t_c \) denotes the true label of the image, and \( \epsilon_a \) represents the attack budget, which constrains the magnitude of the perturbation. This attack can be effectively approximated using the Projected Gradient Descent (PGD) algorithm. PGD iteratively updates the perturbation \( \delta \) by taking steps in the direction of the gradient of the loss function, while projecting the perturbation back onto the feasible set defined by the attack budget \( \epsilon_a \). This iterative process ensures that the perturbation remains within the specified bounds while maximizing the loss function.




\section{Feature Perception Threshold Counterattack Noise}

Existing test-time defense methods for CLIP suffer from several limitations: they employ fixed-intensity counter-noise that fails to adapt to varying attack strengths and image complexities; they cannot reliably distinguish clean from adversarial samples, leading to performance degradation on benign inputs; and they often require manual parameter tuning for different scenarios. To overcome these issues, this section introduces the Feature Perception Threshold Counterattack Noise (FPT-Noise) framework.

\subsection{Dynamic Feature Modulation for Adaptive Noise Intensity}
Adversarial perturbations are inherently subtle disturbances deliberately designed to exploit the decision boundaries of machine learning models. The distribution of these perturbations varies significantly with the specific attack algorithm employed (e.g., BIM, PGD, CW) and the intrinsic content of the targeted images. For instance, images with simpler structures (e.g., handwritten digits from the MNIST dataset) can tolerate lower noise levels due to their lower complexity. In contrast, more complex images (e.g., from ImageNet) often require higher noise intensities to effectively mask adversarial perturbations. Traditional approaches using fixed-intensity noise fail to account for these disparities, limiting their effectiveness across diverse scenarios.We design a Dynamic Feature Modulator (DFM) that operates in a pre-training-free manner. This module quantifies and adapts to the unique features of each image (e.g., texture, edges, semantic structure) to dynamically compute the optimal noise amplitude \(\sigma\) tailored to the specific image.

The DFM is a sophisticated neural module that comprises a multi-head self-attention layer and layer normalization. It processes image features extracted by the CLIP model to capture contextual dependencies within the image content. Specifically, given the image feature \(f_\theta(X)\) from CLIP’s vision encoder \(f_\theta\), the adapted features are derived via the self-attention processing block, formulated as:

\begin{equation}
 \sigma = M(f_\theta(X))
\end{equation}

where \(M\) denotes the Dynamic Feature Modulation operation, and \(\sigma\) represents the dynamically computed noise intensity. This adaptive approach ensures that the noise intensity is precisely calibrated to the specific characteristics of each image, thereby enhancing the robustness of the model against adversarial attacks.

\begin{figure*}[t]
    \includegraphics[width=\linewidth]{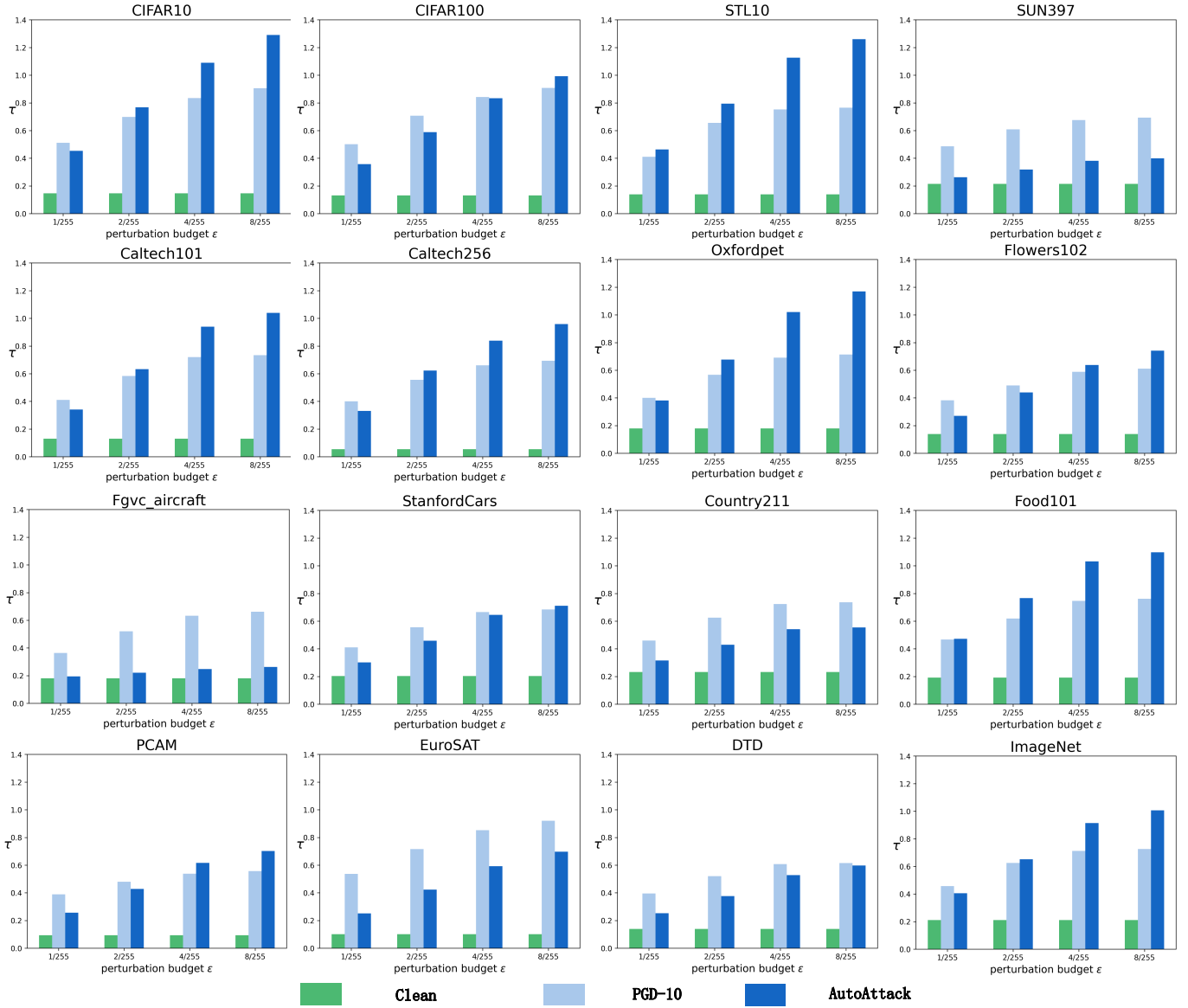}
    \caption{Average $\tau$ across 16 image datasets.}

    \label{fig3}
\end{figure*}

\subsection{Feature Perception Threshold}
In a recent study, Xing et al. \cite{xing2025clip} observed that adversarial images exhibit counterintuitive robustness to small random noise, and are vulnerable only to sufficiently large disturbances. To address this observation, they introduced a stochastic variable \(\tau\) to identify whether an image is "falsely stable." Specifically, when small random noise (\(\epsilon_{\text{random}} = 1/255, 4/255\)) is applied, the ratio of \(L_2\) drift in the latent space is unusually small—indicating the image is trapped in a "toxic region" and rendered "falsely stable" by an adversary. The variable \(\tau\) is defined as:

\begin{equation}
  \tau =  \frac{\| f_\theta(X + \delta_0) - f_\theta(X) \|}{\| f_\theta(X) \|}
\end{equation}

where \(\delta_0\) represents a small random noise perturbation. If \(\tau\) is less than a user-defined threshold \(\tau_{\text{init}}\), a counterattack is initiated; otherwise, a random noise value is returned.

However, our analysis identifies key limitations in this approach: it exhibits limited effectiveness in distinguishing adversarial images from clean ones, particularly for fine-grained images.
Thus, a method is needed to differentiate clean and adversarial images with higher accuracy.



Our experiments reveal a distinct behavioral difference between clean and adversarial images when exposed to noise of varying intensities: clean samples typically exhibit subtle feature shifts, while adversarial images display significantly larger shifts. This observation underpins our proposed Feature Perception Threshold strategy, which measures the difference in an image’s response to two distinct noise levels to enhance detection accuracy.

We define a new metric, Feature Perception Threshold, as follows:

\begin{equation}
  \tau =  \frac{\| f_\theta(X + \delta_1) - f_\theta(X) \| -   \| f_\theta(X + \delta_0) - f_\theta(X) \|}{\| f_\theta(X) \|}
\end{equation}

where \(\delta_0\) is a random noise at \(\epsilon=4/255\), and \(\delta_1\) is a random noise at \(\epsilon=32/255\), which is greater than \(\delta_0\). This metric effectively captures the differential response of the image to varying noise intensities.

We tested our method on 16 datasets used in the Sec \ref{shiyan}, with results shown in Figure \ref{t3}. Our approach effectively distinguishes whether an image has been attacked, demonstrating superior performance compared to the existing method.

\begin{algorithm}[tb]
\caption{FPT-Noise Framework} 
\label{al}
\textbf{Input}: Image $X$, CLIP encoder $f_{\theta}$, stability threshold $\tau_{init}$, random noise $\delta_0,\delta_1$, standard Gaussian noise $\epsilon_{0} \sim \mathcal{N}(0, 1)$  \\
\textbf{Output}: Defended image $X_{def}$  

\begin{algorithmic}[1]

\STATE Extract features $f_{\theta}(X)$
\STATE Compute dynamic noise scale: $\sigma = M(f_{\theta}(X))$
\STATE Compute Feature Perception Threshold  $  \tau =  \frac{\| f_\theta(X + \delta_1) - f_\theta(X) \| -   \| f_\theta(X + \delta_0) - f_\theta(X) \|}{\| f_\theta(X) \|}$
\STATE Compute $k=exp(\tau-\tau_{init})$
\STATE Generate init noise:
\STATE \quad $\delta = k*\sigma *\epsilon_{0}$
\STATE Optimization through loss function: $ \delta_{c} = argmax||f_\theta(X+\delta) - f_\theta(X)||$

\IF{$\tau > \tau_{init}$}
\STATE Apply full counterattack: $X_{def} = X + \delta_{c} $
\ELSE 
\STATE Compute norm ratio: $r =   \| f_\theta(X+\delta_c) \| / \| f_\theta(X) \|$
\IF{$r > \beta$}
\STATE \quad $X_{def} = X +  \delta_{\text{c}} $
\ELSE

\STATE \quad $X_{def} = X + exp((\tau-\tau_{init})\cdot10) \cdot \delta_c $
\ENDIF
\ENDIF
\STATE Ensemble with TTE transformations
\STATE \textbf{return} $X_{def}$
\end{algorithmic}
\end{algorithm}

\subsection{Counterattack Noise}

For the original image \(X\), we generate corresponding noise scales \(\sigma\), which are specifically tailored to the unique characteristics of the image. The Gaussian noise \(\delta\) is then computed as follows:

\begin{equation}
  \delta = k \cdot \sigma \cdot \epsilon_{0}
\end{equation}

Here, \(\epsilon_{0} \sim \mathcal{N}(0, 1)\) denotes standard Gaussian noise, ensuring that the noise introduced is both random and unbiased. The adaptive parameter \(k\) plays a pivotal role in adjusting the counterattack intensity against different types of attacks.
Specifically, \(k\) is defined as:

\begin{equation}
  k = \exp(\tau - \tau_{\text{init}}), \quad \text{with} \quad \min = 1.0, \quad \max = 6.0
\end{equation}

This exponential scaling ensures that the counterattack intensity is dynamically adjusted based on the initial perturbation \(\tau\), relative to a baseline \(\tau_{\text{init}}\). The bounds on \(k\) ensure that the counterattack remains within a reasonable range, thereby preventing excessive perturbation that could degrade the image quality.

To leverage the feature shift induced by initial perturbations and drive adversarial images (which subvert CLIP’s predictions) away from the "toxic" feature regime dominated by adversarial manipulations, we draw inspiration from the work of Schlarmann et al. \cite{schlarmann2024robust} and Xing et al. \cite{xing2025clip}. We adopt an identical loss formulation to construct the Test-Time counterattack perturbation \(\delta_{c}\). Specifically, we treat the embedding of the clean image \(f_\theta({X})\) as a semantic anchor and optimize \(\delta_{c}\) to maximize the \(L_2\) distance between the embedding of the counterattacked image \(f_\theta(X + \delta_{c})\) and this anchor:

\begin{equation}
    \delta_{c} = \arg\max_{\delta} \left\| f_\theta(X + \delta) - f_\theta(X) \right\|_2
\end{equation}

\subsection{Scene-Aware Regulation}

When the feature norm \(\tau\) of an image exceeds the initial threshold \(\tau_{\text{init}}\), the image is classified as adversarial, and a counteracting perturbation \(\delta_{c}\) is introduced to neutralize the adversarial effects. However, certain images may remain below the threshold \(\tau_{\text{init}}\) even after being subjected to adversarial attacks. These include both clean images and potential adversarial samples with minimal perturbations.

To address this scenario, we propose a weighted suppression mechanism to prevent over-perturbation for samples with \(\tau \leq \tau_{\text{init}}\):

\begin{equation}
   W = \exp((\tau- \tau_{\text{init}}) \cdot 10)
\end{equation}

Here, \(W\) represents the suppression weight, which exponentially decays as the difference between \(\tau_{\text{init}}\) and \(\tau\) increases. This mechanism ensures that only necessary counter-perturbations are applied, thereby preserving the integrity of clean images.

\begin{figure*}
    \includegraphics[width=\linewidth]{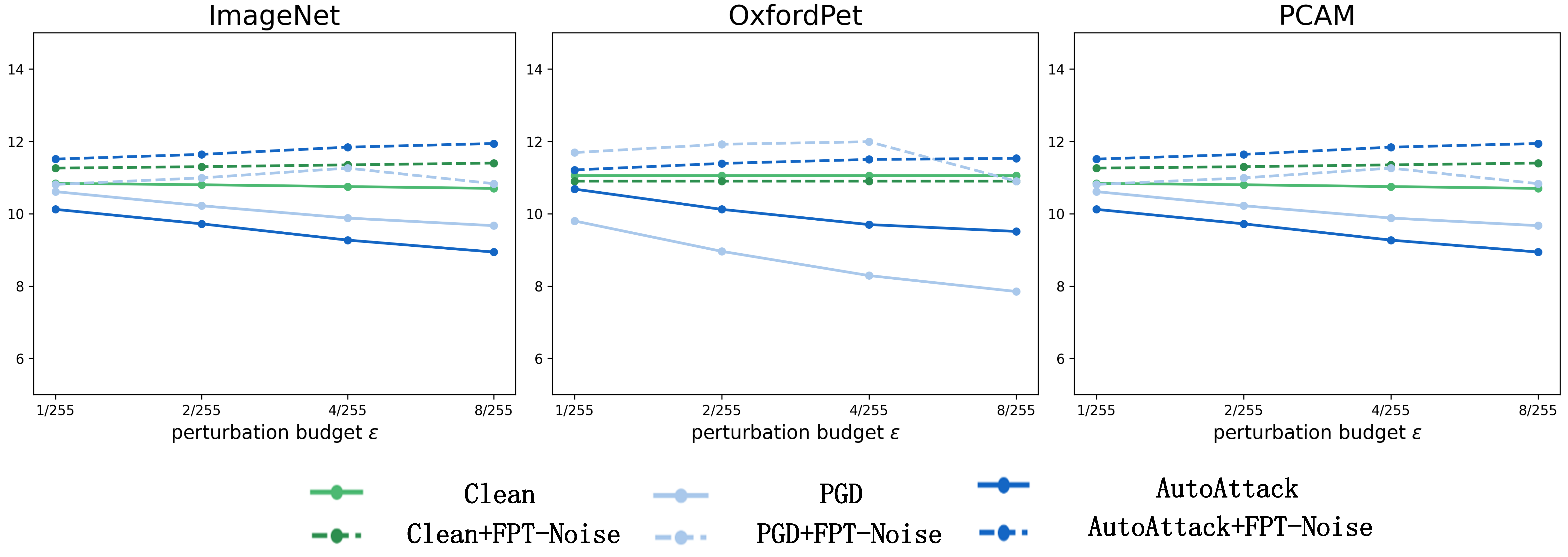}
    \caption{We utilize CLIP to compute the feature norm $||f_\theta(X+\delta_c)||$ of images with added FPT-Noise and the feature norm  $||f_\theta(X)||$ of initial images. We selected three datasets from different domains for analysis.}
    \label{diff}
\end{figure*}
For images that have been attacked, the feature norm tends to decrease, particularly under strong adversarial attacks. By adding counterattack noise, we aim to restore the feature norm magnitude. As shown in Figure \ref{diff}, the feature norm of the clean image remains relatively stable, while more intense attack images cause the feature norm to decrease. By introducing counterattack noise, the feature norm of the attacked image recovers, whereas the clean image exhibits robustness to noise.
This property can be leveraged to distinguish between attacked and clean images below the threshold \(\tau_{\text{init}}\)

We utilize CLIP to compute the feature norm $||f_\theta(X+\delta_c)||$ of images with added FPT-Noise and the feature norm  $||f_\theta(X)||$ of initial images, then calculate their ratio as follows:

\begin{equation}
    r = \frac{||f_\theta(X+\delta_c)||}{||f_\theta(X)||}
\end{equation}

The final counter-perturbation $\delta_{final}$ is determined as follows:

\begin{equation}
    \delta_{\text{final}} =  
    \begin{cases}
        \text{if } \tau >\tau_{init} & \delta_{c}   \\
         \text{else}\\
          \text{if } r>\beta &  \delta_{c} \\
            \text{else} &  W *\delta_{c} \\
    \end{cases}
\end{equation}

Here, \(\beta\) is a hyperparameter. When \(r > \beta\), the feature norm increases after \(\delta_c\) injection, indicating that the image is likely adversarial and the full \(\delta_c\) is retained to neutralize adversarial effects. When \(r \leq \beta\), the feature norm shows little recovery, indicating a clean image, and the suppression weight \(W\) scales down \(\delta_c\) to prevent unnecessary distortion.

After injecting the targeted counterattack noise, we further enhance the robustness of the system by applying Test-Time Transformation Ensembling (TTE) \cite{perez2021enhancing}. This technique involves applying transformations such as flipping and cropping to the input image. These transformations are particularly effective in invalidating high-frequency adversarial perturbations, which are typically viewpoint-dependent. By focusing the model's attention on low-frequency semantic regions, TTE further mitigates residual noise effects and improves the overall robustness of the system.

The entire process is summarized in Algorithm \ref{al}.

\begin{table*}[!tb]  
    \scriptsize
    \caption{Classification accuracy (\%) on adversarial images (Rob.) under 10-step PGD at $\epsilon=1/255$ and clean images (Acc.) averaged on 16 datasets.}
        \renewcommand{\arraystretch}{1.0}  
        \centering  
        \begin{tabular}{@{}ccccccccccccc@{}}  
            \toprule
            \multirow{2}{*}{Dataset} & \multirow{2}{*}{Metric} & \multirow{2}{*}{CLIP} & \multicolumn{3}{c}{Adversarial Fine-tuning} & \multicolumn{4}{c}{Test-Time Defence}  & \multirow{2}{*}{FPT-Noise(ours)} & \multirow{2}{*}{$\Delta$} \\  
            \cmidrule(lr){4-6} \cmidrule(lr){7-10}  
            & & & TeCoA & PMG-AFT & FARE & HD & Anti-adv & AOM & TTC & & & \\  
            \midrule  
            \multirow{2}{*}{CIFAR10} & Rob. & 0.74 & 33.45 & 40.82 & 47.15 & 31.41 & 30.31 & 47.51 & 28.75  & \textbf{75.93} & +75.19 \\  
            & Acc. & \textbf{85.12} & 64.83 & 70.51 & 79.15  & 68.82 & 66.74 & 81.42 & 81.18  & 84.39 & -0.73 \\  
            \midrule  
            \multirow{2}{*}{CIFAR100} & Rob. & 0.26 & 18.72 & 22.79 & 23.60 & 13.86 & 15.14 & 14.96 & 14.31  & \textbf{48.52} & +48.26 \\  
            & Acc. & 57.14 & 36.18 & 40.15 & 47.53 & 38.94 & 38.66 & 53.72 & 56.34  & \textbf{58.17} & +1.03 \\  
            \midrule
            \multirow{2}{*}{STL10} & Rob. & 11.0 & 70.29 & 72.85 & 73.89 & 60.33 & 56.30 & 68.78 & 76.34  & \textbf{94.46} & +83.46 \\  
            & Acc. & 96.40 & 87.18 & 88.73 & 91.35 & 92.05 & 88.06 & 95.02 & \textbf{96.70}  & 95.60 & -0.80 \\  
            \hline  
            \multirow{2}{*}{Caltech101} & Rob. & 14.67 & 55.72 & 60.89 & 59.40 & 48.57 & 55.60 & 60.45 & 65.69  & \textbf{82.56} & +67.89 \\  
            & Acc. & 85.66 & 71.45 & 75.68 & 75.11 & 77.52 & 80.12 & 81.23 & 80.53  & \textbf{86.55} & +0.89 \\  
            \hline  
            \multirow{2}{*}{Caltech256} & Rob. & 8.47 & 43.38 & 45.67 & 52.73 & 38.87 & 44.13 & 49.11 & 60.23  & \textbf{77.02} & +68.55 \\  
            & Acc. & \textbf{81.72} & 78.31 & 61.32 & 72.06 & 71.70 & 74.69 & 77.26 & 79.79  & 80.98 & -0.74 \\  
            \hline  
            \multirow{2}{*}{OxfordPets} & Rob. & 1.04 & 38.01 & 42.05 & 31.07 & 25.92 & 32.00 & 45.95 & 57.24  & \textbf{72.14} & +71.10 \\  
            & Acc. & \textbf{87.44} & 62.33 & 65.67 & 79.37 & 77.38 & 81.47 & 82.98 & 83.35  & 87.00 & -0.44 \\  
            \hline  
            \multirow{2}{*}{Flowers102} & Rob. & 1.14 & 22.15 & 23.28 & 30.22 & 13.50 & 19.27 & 33.32 & 38.98  & \textbf{52.95} & +51.81 \\  
            & Acc. & 65.46 & 36.62 & 37.25 & 47.98 & 50.09 & 57.91 & 62.38 & 64.16  & \textbf{65.95} & +0.49 \\  
            \hline  
            \multirow{2}{*}{FGVC Aircraft} & Rob. & 0.00 & 2.67 & 2.41 & 4.68 & 2.88 & 3.33 & 7.77 & 14.52  & \textbf{18.90} & +18.90 \\  
            & Acc. & \textbf{20.10} & 5.12 & 5.38 & 12.84 & 13.08 & 15.36 & 16.35 & 17.94  & 19.11 & -0.99 \\  
            \hline  
            \multirow{2}{*}{StanfordCars} & Rob. & 0.02 & 8.95 & 11.42 & 15.98 & 5.99 & 9.59 & 14.31 & 33.01  & \textbf{47.16} & +47.14 \\  
            & Acc. & 52.02 & 20.73 & 25.68 & 38.68 & 36.08 & 42.52 & 52.10 & 48.16  & \textbf{52.12} & +0.10 \\  
            \hline  
            \multirow{2}{*}{SUN397} & Rob. & 1.14 & 19.15 & 22.79 & 23.36 & 12.71 & 19.08 & 29.28 & 41.49  & \textbf{59.48} & +58.34 \\  
            & Acc. & 58.50 & 36.88 & 37.76 & 38.85 & 39.69 & 49.56 & \textbf{58.66} & 55.73  & 57.89 & -0.61 \\  
            \hline  
            \multirow{2}{*}{Country211} & Rob. & 0.04 & 1.96 & 2.03 & 3.85 & 1.42 & 2.09 & 3.79 & 7.21  & \textbf{12.04} & +12.00 \\  
            & Acc. & \textbf{15.25} & 4.58 & 4.82 & 9.26 & 7.53 & 9.67 & 12.95 & 13.26  & 13.13 & -2.12 \\  
            \hline  
            \multirow{2}{*}{Food101} & Rob. & 0.70 & 14.12 & 18.35 & 28.65 & 16.67 & 23.75 & 19.35 & 57.91  & \textbf{73.18} & +72.48 \\  
            & Acc. & \textbf{83.88} & 30.15 & 36.87 & 55.31 & 58.33 & 64.21 & 76.84 & 82.18  & 80.21 & -3.67 \\  
            \hline  
            \multirow{2}{*}{EuroSAT} & Rob. & 0.03 & 12.18 & 12.42 & 10.67 & 12.96 & 10.99 & 6.73 & 9.39  & \textbf{25.28} & +25.25 \\  
            & Acc. & 38.25 & 16.37 & 18.72 & 21.88 & 29.26 & 24.40 & 32.53 & \textbf{51.81}  & 38.02 & -0.23 \\
            \hline  
            \multirow{2}{*}{DTD} & Rob. & 2.98 & 17.42 & 15.18 & 15.64 & 13.62 & 15.69 & 19.10 & 27.98  & \textbf{37.23} & +34.25 \\  
            & Acc. & \textbf{40.64} & 25.35 & 21.59 & 32.07 & 30.27 & 33.14 & 38.01 & 36.98  & 39.15 & -1.49 \\  
            \hline  
            \multirow{2}{*}{PCAM} & Rob. & 0.08 & 48.03 & 46.37 & 20.23 & 13.62 & 29.12 & 50.00 & 52.85  & \textbf{51.68} & +51.60 \\  
            & Acc. & 52.02 & 50.17 & 49.85 & 50.54 & 48.79 & 48.55 & 49.85 & \textbf{53.07}  & 51.17 & -0.85 \\
		\hline  
    	\multirow{2}{*}{ImageNet} & Rob. & 1.15 & 19.07 & 21.25 & 24.00 & 14.38 & 19.05 & 22.37 & 38.41  & \textbf{53.18} & +52.03 \\  
			& Acc. & \textbf{59.69} & 34.68 & 36.31 & 40.79 & 44.25 & 49.57 & 59.05 & 49.39  & 57.23 & -2.46 \\
			\hline  
			\multirow{2}{*}{Avg.} & Rob. & 2.70 & 26.38 & 28.61 & 29.07 & 20.42 & 24.09 & 27.11& 39.17  & \textbf{55.11} & +52.41 \\
			& Acc. & \textbf{61.51} & 40.42 & 42.15 & 49.55 & 48.99 & 51.52 &58.27 & 59.75  & 60.39 & -1.12 \\
			\bottomrule
		\end{tabular}

    \label{t1}  
	\begin{flushleft}
		\footnotesize \textit{Note:} $\Delta$ shows the improvement of FPT-Noise over CLIP baseline. FPT-Noise(ours) column uses TTC data from logs. \textbf{Bold values} indicate the maximum accuracy in each row.
	\end{flushleft}
\end{table*}

\section{Experience}
\label{shiyan}
\subsection{Experimental Setup}
To validate the effectiveness of the FPT-Noise framework in improving the zero-shot adversarial robustness of CLIP, we conducted extensive experiments on multiple benchmark datasets by ViT-B-32. We set $\beta$ to 1.125, and $\tau_{init}$ to 0.32. All experiments were conducted on NVIDIA-V100s. Detailed experiments on hyperparameters are provided in the appendix.

\subsection{Datasets}
This study builds on previous work that examined the adversarial robustness of CLIP \cite{schlarmann2024robust,xing2025clip}. We selected 16 representative datasets, including CIFAR10 \cite{cifar}, CIFAR100 \cite{cifar}, STL10 \cite{STL}, Caltech101 \cite{caltech101}, Caltech256 \cite{caltech256}, OxfordPets \cite{oxfordpets}, Flowers102 \cite{flower}, Food101 \cite{food}, StanfordCars \cite{stanfordcars}, SUN397 \cite{sun}, Country211 \cite{clip}, FGVCAircraft \cite{FGVC}, EuroSAT \cite{eurosat}, DTD \cite{dtd}, PCAM \cite{pcam}, and ImageNet \cite{imagenet}. These datasets cover a variety of image classification tasks, enabling a comprehensive evaluation of the model's robustness.

\begin{table*}[t]  
    \caption{Classification accuracy (\%) on adversarial images (Rob.) under Autoattack at $\epsilon=1/255$ and clean images (Acc.) averaged on 16 datasets.}  
    \label{aa}  
    \scriptsize
    \renewcommand{\arraystretch}{1.0}  
        \centering  
        \begin{tabular}{@{}ccccccccccccc@{}}  
            \toprule  
            \multirow{2}{*}{Dataset} & \multirow{2}{*}{Metric} & \multirow{2}{*}{CLIP} & \multicolumn{3}{c}{Adversarial Fine-tuning} & \multicolumn{4}{c}{Test-Time Defence}  & \multirow{2}{*}{FPT-Noise(ours)} & \multirow{2}{*}{$\Delta$} \\  
            \cmidrule(lr){4-6} \cmidrule(lr){7-9}  
            & & & TeCoA & PMG-AFT & FARE & HD & Anti-adv & AOM & TTC & & & \\  
            \midrule  
            \multirow{2}{*}{CIFAR10} & Rob. & 0.03 & 30.53 & 37.37 & 45.32 & 0.03 & 39.40 & 49.00 & 4.04  & \textbf{76.90} & +76.87 \\  
            & Acc. & \textbf{85.05} & 64.61 & 70.69 & 79.15  & 85.05 & 66.87 & 81.45 & 81.37  & 84.15 & -0.90 \\  
            \midrule  
            \multirow{2}{*}{CIFAR100} & Rob. & 0.05 & 16.54 & 19.83 & 21.19 & 0.05 & 22.40 & 15.58 & 8.30  & \textbf{48.57} & +48.52 \\  
            & Acc. & 57.17 & 35.96 & 40.32 & 47.53 & 57.17 & 38.38 &56.32 & 56.36  & \textbf{58.54} & +1.37 \\  
            \midrule  
            \multirow{2}{*}{STL10} & Rob. & 0.00 & 65.53 & 70.27 & 73.02 & 0.00 & 62.85 & 69.25 & 4.33  & \textbf{95.60} & +95.60 \\  
            & Acc. & \textbf{96.40} & 87.40 & 88.56 & 91.35 & 96.40 & 87.86 &95.10 & 95.78  & 95.41 & -0.99 \\  
            \hline  
            \multirow{2}{*}{Caltech101} & Rob. & 0.30 & 51.02 & 57.18 & 59.40 & 0.30 & 62.42 & 61.39 & 9.97  & \textbf{84.55} & +84.25 \\  
            & Acc. & 86.23 & 71.68 & 75.45 & 75.11 & 86.23 & 80.00 &81.35 & \textbf{86.45}  & 86.27 & +0.04 \\  
            \hline  
            \multirow{2}{*}{Caltech256} & Rob. & 0.06 & 40.12 & 45.91 & 50.98 & 0.06 & 52.52 & 49.94 & 7.04  & \textbf{79.32} & +79.26 \\  
            & Acc. & \textbf{82.05} & 78.53 & 61.14 & 72.06 & 82.05 & 74.71 &77.12 & 79.82  & 80.97 & -1.08 \\  
            \hline  
            \multirow{2}{*}{OxfordPets} & Rob. & 0.03 & 34.27 & 36.68 & 31.07 & 0.03 & 40.26 & 47.07 & 6.19  & \textbf{81.66} & +81.63 \\  
            & Acc. & 87.30 & 62.12 & 65.88 & 79.37 & 87.30 & 81.38 &82.82 & 82.94  & \textbf{87.90} & +0.60 \\  
            \hline  
            \multirow{2}{*}{Flowers102} & Rob. & 0.05 & 17.96 & 21.34 & 30.22 & 0.05 & 23.99 & 33.96 & 9.20  & \textbf{55.81} & +55.76 \\  
            & Acc. & \textbf{65.59} & 36.80 & 37.00 & 47.98 & 65.59 & 57.90 &62.33 & 64.25  & 65.41 & -0.18 \\  
            \hline  
            \multirow{2}{*}{FGVC Aircraft} & Rob. & 0.03 & 0.49 & 1.22 & 4.68 & 0.03 & 8.01 & 8.28 & 2.97  & \textbf{18.06} & +18.03 \\  
            & Acc. & \textbf{20.07} & 5.31 & 5.55 & 12.84 & 20.07 & 15.48 &16.30 & 18.63  & 18.81 & -1.26 \\  
            \hline  
            \multirow{2}{*}{StanfordCars} & Rob. & 0.02 & 5.67 & 9.64 & 15.98 & 0.02 & 16.30 & 15.31 & 4.70  & \textbf{48.58} & +48.56 \\  
            & Acc. & 52.02 & 20.91 & 25.44 & 38.68 & 52.02 & 42.47 &52.07 & 48.35  & \textbf{52.27} & +0.25 \\  
            \hline  
            \multirow{2}{*}{SUN397} & Rob. & 0.06 & 15.30 & 20.55 & 23.36 & 0.06 & 27.03 &29.40 & 6.68  & \textbf{57.96} & +57.90 \\  
            & Acc. & \textbf{58.86} & 36.69 & 37.98 & 38.85 & 58.86 & 49.51 &58.71 & 55.88  & 58.37 & -0.49 \\  
            \hline  
            \multirow{2}{*}{Country211} & Rob. & 0.01 & 0.52 & 1.15 & 3.85 & 0.01 & 4.87 & 29.20 & 2.28  & \textbf{12.30} & +12.29 \\  
            & Acc. & \textbf{15.21} & 4.75 & 4.64 & 9.26 & 15.21 & 9.67 &12.90 & 13.07  & 13.42 & -1.79 \\  
            \hline  
            \multirow{2}{*}{Food101} & Rob. & 0.04 & 10.42 & 16.75 & 28.65 & 0.04 & 27.83 & 19.58 & 4.51  & \textbf{80.19} & +80.15 \\  
            & Acc. & \textbf{83.82} & 29.98 & 36.61 & 55.31 & 83.82 & 64.03 & 76.84& 82.25  & 80.90 & -2.92 \\  
            \hline  
            \multirow{2}{*}{EuroSAT} & Rob. & 0.09 & 9.68 & 10.69 & 10.67 & 0.09 & 15.65 & 19.35 & 18.15  & \textbf{27.84} & +27.75 \\  
            & Acc. & 38.25 & 16.58 & 18.53 & 21.88 & 38.25 & 24.23 & 38.04 & \textbf{51.60}  & 39.25 & +0.99 \\  
            \hline  
            \multirow{2}{*}{DTD} & Rob. & 0.05 & 14.16 & 12.52 & 15.64 & 0.05 & 20.80 & 21.85 & 7.39  & \textbf{35.85} & +35.80 \\  
            & Acc. & \textbf{40.16} & 25.16 & 21.76 & 32.07 & 40.16 & 32.50 &37.92 & 36.91  & 39.20 & -0.96 \\  
            \hline  
            \multirow{2}{*}{PCAM} & Rob. & 0.21 & 46.42 & 44.82 & 20.23 & 0.21 & 41.49 & 41.02 & 8.15  & \textbf{50.55} & +50.34 \\  
            & Acc. & \textbf{52.80} & 49.96 & 50.03 & 50.54 & 52.80 & 48.50 &49.97 & 52.94  & 50.04 & -2.76 \\  
            \hline  
            \multirow{2}{*}{ImageNet} & Rob. & 0.04 & 14.79 & 18.84 & 23.28 & 0.04 & 26.46 & 23.73 & 6.26  & \textbf{56.02} & +55.98 \\  
            & Acc. & \textbf{59.73} & 34.89 & 36.12 & 40.79 & 59.73 & 49.51 &59.01 & 49.67  & 57.42 & -2.31 \\  
            \hline  
            \multirow{2}{*}{Avg.} & Rob. & 0.07 & 23.34 & 26.55 & 28.60 & 0.07 & 30.77 &32.31 & 6.89  & \textbf{56.86} & +56.79 \\  
            & Acc. & \textbf{61.29} & 40.25 & 42.30 & 49.55 & 61.29 & 51.44 & 58.33& 59.77  & 60.52 & -0.77 \\  
            \bottomrule  
        \end{tabular}  

    \begin{flushleft}  
        \footnotesize \textit{Note:} $\Delta$ shows the improvement of FPT-Noise over CLIP baseline. FPT-Noise(ours) column uses TTC data from logs. \textbf{Bold values} indicate the maximum accuracy in each row.  
    \end{flushleft}  
\end{table*}

\subsection{Baseline}
Given the scarcity of dedicated Test-Time defense methods for CLIP in the existing literature, we have adapted several established adversarial robustness techniques that do not rely on auxiliary networks. Our comparative analysis focuses on two particularly relevant approaches: Test-Time Defense and Adversarial Fine-tuning.
\subsubsection{Test-Time defense}

Hedge Defense (HD): Following  \cite{wu2021attacking}, we implemented their cross-entropy maximization approach across all candidate classes. This method capitalizes on the observation of smoother loss landscapes near ground-truth categories. For our CLIP adaptation, we maintained the original paper's configuration of 20 iterative steps with $4/255$.

Anti-adversary Defense (Anti-adv): Building on  \cite{anti}, we modified their confidence-reinforcement strategy for CLIP by maximizing the cosine similarity between adversarial images and their most probable text embeddings. This implementation utilized a perturbation budget of $4/255$  across two optimization steps.

Anchor-guided One-step linear Movement (AOM): we reproduced the results based on the paper's content, with hyperparameters consistent with those specified in the paper.

$\tau$-thresholded Weighted Counterattacks (TTC): Based on the settings in the original paper \cite{xing2025clip}, we set the number of steps to 2 when the attack budget epsilon is $1/255$, and set the number of steps to 5 when the attack budget epsilon is $4/255$.

\subsubsection{Adversarial Fine-tuning}

We implemented three state-of-the-art methods: TeCoA, PMG-AFT, and FARE, corresponding to the works of \cite{mao2022understanding},  \cite{wang2024pre}, and \cite{schlarmann2024robust}, respectively. All fine-tuning procedures were conducted on TinyImageNet, using 2-step PGD attacks with a perturbation magnitude of 1/255 and a learning rate of $5 \times 10^{-5}$. These implementations followed the established protocols outlined in the referenced literature.


\subsection{Adversarial attack methods}
To assess the model's adversarial robustness, several prevalent adversarial attack methods are utilized, including PGD, AutoAttack. The standard version of AutoAttack is employed. 

\subsection{Quantitative Robustness Evaluation of CLIP}

We evaluate the robustness of all methods under the PGD-10 attack with an attack budget of $\epsilon$ = 1/255. This evaluation scenario assumes that attackers have full access to the model's weights and gradients during deployment but are unaware of any Test-Time defense operations implemented by end users. The accuracy results for both adversarial images and clean images are presented in Table \ref{t1}. 

It should be noted that these fine-tuned methods were trained on the Tiny-ImageNet dataset following the original papers.

As shown in Table \ref{t1}, existing methods undermine CLIP's  zero-shot generalizability to some degree. Among Test-Time defenses, TTE's image transformations yield limited robustness gains. Anti-adv and TTC, like ours, use architectural designs for noise counteraction, but our FPT-Noise stands out: by leveraging CLIP's visual encoder to generate dynamic, frequency-adaptive image-specific noise, it boosts robust accuracy by +52.41\% on average vs. original CLIP, with only a marginal -1.12\% clean accuracy drop, proving stable as an test-time defense. Notably, this defense strategy can be considered a trade-off, as it almost does not compromise the model's strong recognition capabilities.

We also evaluated the effectiveness of our approach in defending against AutoAttack in Table \ref{aa}. Since AutoAttack has a greater impact on feature norm, the Scene-Aware Regulation module we designed demonstrates more pronounced perception and delivers more significant defense effectiveness.

We further evaluated the method against PGD attacks with varying perturbation budgets: $\epsilon$ = 1/255, 2/255, 4/255, 8/255. Following the setup in the TTC paper, we set the counterattack steps to 5 for the \(\epsilon=4/255\) scenario. As demonstrated in Figure \ref{fig1}(a), our method effectively generates robust counterattack noise across different attack types and intensities, while requiring only 2 counterattack step—eliminating the need for scenario-specific parameter tuning.

\subsection{FPT-Noise on Finetune CLIP}
In order to provide further validation of the generalizability of our approach, we applied the FPT-Noise method to fine-tuning-based methods, including TeCoA, PMG-AFT, and FARE, in conjunction with TTC for the purpose of comparison. As demonstrated in Table \ref{t2}, FPT-Noise marginally enhances the performance of TeCoA, PMG-AFT, and FARE.

Fine-tuning enhances the robustness of the model, but it comes at the expense of the model's original performance. The efficacy of our FPT-Noise is contingent upon CLIP's capacity to discern variations within the feature space, thereby precluding substantial enhancement through fine-tuning methodologies. Adversarial fine-tuning is employed by methods such as TeCoa and PMG-AFT, while FARE utilizes an unsupervised approach to fine-tune CLIP, thereby demonstrating that unsupervised fine-tuning exerts a lesser influence on the CLIP model's original performance.


\begin{table}[t]
    \centering

    \scriptsize
    \renewcommand{\arraystretch}{1.0}  
    \setlength{\tabcolsep}{4pt}  
    \caption{FPT-Noise employed on adversarially finetuned models at test time. Classification accuracy (\%) on adversarial images (Rob.) under 10-step PGD at $\epsilon$ = 1/255 and clean images (Acc.) averaged on 16 datasets. } 

    \begin{tabular}{c|cc}
    
    \toprule 
    \textbf{(\%)} & \textbf{Rob.} & \textbf{Acc.}  \\
    \midrule 
    TeCoA & 26.38 & 40.42 \\
    TeCoA + TTC & 29.06 & 39.81 \\
    \midrule
   \textbf{ TeCoA + FPT-Noise} & 29.32 & 39.92 \\
    \midrule
    PMG-AFT & 28.61 & 42.15 \\
    PMG-AFT + TTC & 30.81 & 41.89 \\
    \midrule
    \textbf{PMG-AFT + FPT-Noise} & 31.83 & 41.95  \\
    \midrule
    FARE & 29.07 & 49.55 \\
    FARE + TTC & 32.85 & 48.12 \\
    \midrule
    \textbf{FARE + FPT-Noise} & 35.41 & 48.91 \\
    \bottomrule 
    \end{tabular}

    \label{t2}
\end{table}

\begin{table}[htbp]
\centering
    \renewcommand{\arraystretch}{1.0}  
    \setlength{\tabcolsep}{4pt}  
\caption{The present study aims to assess the impact of various modules. Classification accuracy (\%) on adversarial images (Rob.) under 10-step PGD and AutoAttack at $\epsilon$ = 1/255 and clean images (Acc.) averaged on 16 datasets. } 
\begin{tabular}{ccc|cccc}
\hline
    \multicolumn{3}{c}{\textbf{Module}} &\multicolumn{2}{c}{PGD}&\multicolumn{2}{c}{AutoAttack}\\
\hline
 DFM. & Feature. & Scene. & \textbf{Rob.} & \textbf{Acc.} & \textbf{Rob.} & \textbf{Acc.}\\

 \hline
$ \checkmark$ & $\checkmark$ & $\times$&53.37&60.68&49.68&60.55\\
$ \checkmark$ & $\times$ & $\checkmark$&46.72 &58.43&10.38&58.67\\
$ \times$ & $ \checkmark$ & $\checkmark$&20.40 &61.14&28.77&61.11\\
$ \checkmark$ & $ \checkmark$ & $\checkmark$&55.11 &60.39&56.86&60.52\\
\hline
\end{tabular}

\label{t3}
\end{table}

\begin{table}[htbp]
    \centering
    \scriptsize
    \caption{We tested different versions of the CLIP. Classification accuracy (\%) on adversarial images (Rob.) under 10-step PGD and AutoAttack at $\epsilon$ = 1/255 and clean images (Acc.) averaged on 16 datasets. } 
    \renewcommand{\arraystretch}{1.0}  
    \setlength{\tabcolsep}{4pt}  
    \begin{tabular}{c|ccccc}
    \toprule 
    \textbf{Method} &&\multicolumn{2}{c}{PGD}&\multicolumn{2}{c}{AutoAttack}\\
    \hline
    (\%) &Metric & \textbf{Rob.} & \textbf{Acc.} & \textbf{Rob.} & \textbf{Acc.}\\
     \hline
     \multirow{2}{*}{ViT-B-32}  &Clean &2.70&61.51 &0.07&61.29\\
     &\textbf{Ours}& 55.11 & 60.30&56.86&60.52 \\
     \hline
     \multirow{2}{*}{ViT-B-16} &Clean &4.02&65.56 & 0.10&61.29\\
     &\textbf{Ours}& 60.24& 63.12&62.10&63.60\\
     \hline
     \multirow{2}{*}{ViT-L-14} &Clean &6.25 &71.89  &0.10 &71.83\\
     &\textbf{Ours} &67.96&70.28 &68.52 & 70.59 \\
     \hline
     \multirow{2}{*}{ResNet50} &Clean   & 3.15 & 56.48   & 0.20 & 56.39  \\
     & \textbf{Ours}    & 57.36 & 55.20    & 52.72   & 55.22   \\
     \hline
     \multirow{2}{*}{ResNet101} &Clean    &3.28 &57.68  &0.20    &57.63  \\
   &\textbf{Ours}   &54.81 & 57.44    & 53.15 & 57.36\\
     
    \bottomrule 
    \end{tabular}

    \label{t4}
\end{table}

\subsection{Ablation studies}

\subsubsection{Independent Evaluation of Proposed Method}
To validate the efficacy of our approach, we conducted targeted assessments of three core modules. The following terms are relevant to the present study: Dynamic Feature Modulation, Feature Perception Threshold, Scene-Aware Regulation. We conducted three sets of experiments: (1). We replaced the noise generated by DFM with random noise; (2). We substituted feature perception threshold with the threshold value of TTC\cite{xing2025clip}; (3). We omitted the scene component, similar to TTC, returning only random noise for images failing to meet the threshold condition.

As demonstrated in Table \ref{t3}, the experimental findings substantiate that each designed module fulfills an indispensable role in the framework's performance.

\subsubsection{Comparative Analysis of Model Architectures}
In this paper, previous experiments were based on \textbf{ViT-B-32}.  To further verify the generalizability of our method across CLIP variants, we evaluated it on additional framework configurations. As presented in Table \ref{t4}, the ViT-B-16 and ViT-L-14 architectures—capable of capturing more fine-grained visual features—demonstrate enhanced performance with our approach. Notably, ViT-L-14, with its larger parameter scale and more powerful feature extraction capability, yields the most significant improvements, underscoring the compatibility of our framework with high-capacity CLIP backbones. We also conducted experiments on different architectures, including ResNet50 and ResNet101, where our method similarly achieved excellent performance.

\subsection{Processing Time Consumption Comparison}
We compared processing times with the latest methods such as AOM and TTC. We calculated the processing time required for a single image, with results shown in Table \ref{time}.

\begin{table}[]
    \centering
    \renewcommand{\arraystretch}{1.0}  
    \setlength{\tabcolsep}{4pt}  
    \label{time}
    \caption{Comparison of single image processing time with AOM, TTC}
    \begin{tabular}{c|c|c}
    \toprule 
    \textbf{AOM} & \textbf{TTC} & \textbf{FPT-Noise(Ours)}  \\
    \hline

     35ms   & \textbf{26ms} & 44ms\\

    \bottomrule 
    \end{tabular}

\end{table}

\section{Conclusion}


In this study, we propose FPT-Noise, a novel test-time defense framework to enhance CLIP’s adversarial robustness without costly fine-tuning. Its core contributions are threefold: (1) a Dynamic Feature Modulator generating image-specific and attack-adaptive noise intensity; (2) a Feature Perception Threshold distinguishing clean from adversarial images; (3) Scene-Aware Regulation (guided by a stability threshold) integrated with Test-Time Transformation Ensembling (TTE) to mitigate residual noise and preserve image integrity. Extensive experiments across datasets show FPT-Noise outperforms existing test-time defenses: under AutoAttack, it raises average robust accuracy from 0.07\% to 56.86\% while maintaining high clean-image performance (only a –1.1\% drop). This approach balances robustness and CLIP’s zero-shot generalizability, serving as a practical solution. Future work will optimize dynamic noise generation, extend the framework to other large vision-language models, and integrate it with other defenses for broader adversarial resilience.


\bibliographystyle{IEEEtran}
\bibliography{tmm.bib}

\vfill

\end{document}